\documentclass[
 aps,
 prb,
 reprint,
 superscriptaddress,
 longbibliography,
 nofootinbib
]{revtex4-2}

\usepackage{amsmath}
\usepackage{amssymb}
\usepackage{bm}
\usepackage{graphicx}
\usepackage{hyperref}
\hypersetup{hidelinks}

\begin{document}

\title{Spatial interference changes the symmetry class of a Josephson element}

\author{Takayuki Kubo}
\email{kubotaka@post.kek.jp}
\affiliation{High Energy Accelerator Research Organization (KEK), Tsukuba, Ibaraki 305-0801, Japan}
\affiliation{The Graduate University for Advanced Studies, SOKENDAI, Hayama, Kanagawa 240-0193, Japan}


\begin{abstract}
Spatial interference can break the inversion symmetry of an
effective Josephson potential even in a laterally homogeneous
junction with inversion-symmetric local energy.
We derive a general criterion based on harmonic-dependent
complex spatial form factors: a single local harmonic remains
a shifted cosine, whereas multiple harmonics can acquire phases
incompatible with any common inversion center, permitting cubic
interactions and three-wave mixing.
Phase reflection reverses all odd-order expansion coefficients
about corresponding minima while preserving the even-order ones.
For a capacitively shunted junction at fixed charging energy
and arbitrary offset charge, it also preserves the complete
quantum spectrum, enabling reversal of the odd nonlinear sector
without spectral retuning.
A centered cross junction with multiple local harmonics is one concrete example. Within the imposed-field model, a three-component magnetic field can generate an inversion-asymmetric effective potential, and reversing an odd number of field components exactly reflects it without changing the quantum spectrum of the capacitively shunted junction.
\end{abstract}

\maketitle


\textit{Introduction.---}
The symmetry of a circuit element determines which nonlinear interactions are allowed. For a Josephson element, inversion symmetry about equilibrium forbids odd-order nonlinearities, while breaking it permits a cubic interaction and hence three-wave mixing for phase-preserving amplification, frequency conversion, squeezing, and drive-activated interactions~\cite{Bergeal2010,Frattini2017,Sivak2019,Chapman2023}.

Inversion asymmetry is commonly engineered using multi-junction circuits or symmetry-broken weak links~\cite{Frattini2017,Sivak2019,SchradeFatemi2024,RezaFrolov2026}. Multi-junction interferometers also permit Fourier engineering of the energy--phase relation~\cite{Bozkurt2023}. Previous studies have examined magnetic interference in cross junctions~\cite{Miller1985,Monaco2009,Mawatari2022,Haraoka2024}, anharmonic current--phase relations~\cite{Barash2010}, nonreciprocity in asymmetric junctions~\cite{Hovhannisyan2025}, and field-controlled interference symmetries~\cite{Tsai2026}. This raises a more basic question: can a single, laterally homogeneous, locally reciprocal Josephson junction acquire an inversion-asymmetric effective potential?

Here we show that it can, through spatial interference. We develop a general criterion for this symmetry change and establish how phase reflection reverses odd nonlinearities while preserving the quantum spectrum of a capacitively shunted junction at arbitrary offset charge. As a concrete example, we consider a homogeneous cross junction, where reversing an odd number of magnetic-field components exactly reflects the potential within the imposed-field model. We also identify conditions for a phase-sensitive test distinguishing genuine local Josephson harmonics~\cite{Willsch2024} from circuit-induced ones~\cite{Kim2026}.


\textit{General formulation.---}
We first formulate the symmetry mechanism for a laterally homogeneous, spatially extended reciprocal Josephson element. At zero field, we assume that the stable equilibrium phase coincides with an inversion center of the local Josephson energy, so that all odd-order nonlinearities vanish there. We then apply a magnetic field and follow the resulting state as a spatial phase texture develops across the junction.
We assume a vortex-free two-dimensional weak link, a negligible Josephson self-field, and sufficiently weak magnetic screening that the imposed junction phase texture can be used. Magnetic modification of the local current--phase relation itself is neglected.
The gauge-invariant phase difference across the junction is given by
\begin{eqnarray}
\theta(x,y)=\phi+\delta(x,y),
\end{eqnarray}
where $\phi$ is the spatially uniform circuit phase, while $\delta(x,y)$ describes the spatial variation of the phase across the junction.
For a reciprocal weak link, the local current--phase relation is odd in phase, and therefore the local Josephson energy density $u(\theta)$ is even~\cite{Golubov2004}. It can thus be expanded in cosine harmonics with real coefficients $u_m$ and written generally as
\begin{eqnarray}
u(\theta)=u_0+2\sum_{m\geq1}u_m\cos(m\theta).
\end{eqnarray}
The total Josephson energy is
$U(\phi)=\int_{A_J} u[\theta(x,y)]\,dx\,dy$,
where $A_J$ is the junction area. Substituting the harmonic expansion of
$u(\theta)$ and using $\theta=\phi+\delta(x,y)$, we obtain
\begin{eqnarray}
U(\phi)
&=& A_J u_0
+2A_J\sum_{m\geq1}u_m|F_m|
\cos(m\phi+\alpha_m),
\label{eq:U1}\\
F_m
&:=& \left\langle e^{im\delta}\right\rangle
=\int_{A_J}\frac{dx\,dy}{A_J}\,
e^{im\delta(x,y)}
=|F_m|e^{i\alpha_m},
\label{eq:Fm1}
\end{eqnarray}
where $\alpha_m=\arg F_m$.
The total supercurrent is then given by
$I(\phi)=(2\pi/\Phi_0)dU/d\phi$.

We now ask when the resulting potential remains inversion symmetric, using the phase-inversion-center criterion also relevant to Josephson diodes~\cite{Nikolic2026}. If $\phi_c$ is an inversion center, the potential must satisfy $U(\phi_c+\vartheta)=U(\phi_c-\vartheta)$ for arbitrary $\vartheta$. For the $m$th harmonic,
$\cos[m(\phi_c+\vartheta)+\alpha_m]-\cos[m(\phi_c-\vartheta)+\alpha_m]
=-2\sin(m\phi_c+\alpha_m)\sin(m\vartheta)$.
Thus the $m$th harmonic is symmetric about $\phi_c$ if and only if $\sin(m\phi_c+\alpha_m)=0$.
The full potential is inversion symmetric only if the same $\phi_c$ satisfies this condition for every harmonic $m$ with $u_m|F_m|\neq0$.

A single harmonic always satisfies this condition, since its phase can be removed by shifting the phase origin. When $u_1F_1\neq0$, taking the fundamental as the phase reference, $\phi\to\phi-\alpha_1$ removes its phase shift, and the $m$th harmonic becomes proportional to $\cos(m\phi+\chi_m)$, where $\chi_m=\alpha_m-m\alpha_1$. Thus, $\chi_m$ is the phase-origin-independent mismatch between the $m$th harmonic and the fundamental. The full potential is inversion symmetric about $\phi=0$ if and only if $\chi_m=0$ (${\rm mod}\ \pi$) for every harmonic $m$ with $u_m|F_m|\neq0$. With multiple harmonics, spatial interference generally changes the effective inversion symmetry because their phase shifts cannot all be removed by a single shift of the circuit phase.

To gain analytic insight into the symmetry-breaking mechanism, we now consider the weak-texture limit. The exact symmetry relations and finite-field numerical results below do not rely on this expansion. The interference factor can be expressed in terms of the cumulants $\kappa_n$ of the spatial distribution of $\delta$ as $\ln F_m=\ln \langle e^{im\delta}\rangle =\sum_{n\geq1}\{ (im)^n/n! \}\kappa_n$.
Here the first three cumulants are $\kappa_1=\langle\delta\rangle$, $\kappa_2=\langle(\delta-\langle\delta\rangle)^2\rangle$, and $\kappa_3=\langle(\delta-\langle\delta\rangle)^3\rangle$. Thus, $\kappa_1$ is the mean phase shift, $\kappa_2$ measures the spread of the spatial phase texture, and $\kappa_3$ measures its asymmetry about the mean. For $|m\delta|\ll1$ for the harmonics of interest, the interference phase can be expanded as $\alpha_m=m\kappa_1-(m^3/6)\kappa_3+(m^5/120)\kappa_5+\cdots$.
Even cumulants modify only the magnitude $|F_m|$, whereas the interference phase $\alpha_m$ depends only on odd cumulants.
Using $\chi_m=\alpha_m-m\alpha_1$, we obtain $\chi_m=-(m^3-m)\kappa_3/6+(m^5-m)\kappa_5/120+\cdots$.
The term proportional to $m\kappa_1$ cancels exactly in $\chi_m$, because it corresponds only to a common shift of the phase origin and cannot change the inversion symmetry. The leading relative phase mismatch between different harmonics is therefore proportional to $\kappa_3$, which measures the asymmetry of the spatial phase texture about its mean. In the weak-texture limit, this asymmetry can thus break the effective inversion symmetry when more than one harmonic is present.

To connect this symmetry breaking to the nonlinear response of the Josephson element, we expand the effective potential about a stable minimum $\phi_0$ as $U(\phi_0+q)=U(\phi_0)+\sum_{k\geq2}c_kq^k/k!$, where $c_k=(d^kU/d\phi^k)_{\phi_0}$. Here $c_2$ is the linear restoring curvature, while $c_3$ is the leading inversion-odd nonlinearity.
We choose the phase origin so that the zero-field stable minimum lies at $\phi=0$, assume $u''(0)>0$ and sufficient local smoothness, and define $S_k:=\sum_{m\geq1}u_m m^k$ when termwise differentiation is valid. Expanding $U'(\phi_0)=0$ gives $\phi_0=-\kappa_1-\kappa_3u^{(4)}(0)/[6u''(0)]+O(\delta^5)$. Substituting this shift into $c_3=U'''(\phi_0)$ yields
\begin{eqnarray}
c_3
&=&\frac{A_J \kappa_3}{3}
\left(\frac{S_4^2}{S_2}-S_6\right)
+O(\delta^5)\nonumber\\
&=&\frac{A_J\kappa_3}{6}
\left[u^{(6)}(0)-\frac{[u^{(4)}(0)]^2}{u''(0)}\right]
+O(\delta^5).
\label{eq:c3weak}
\end{eqnarray}
The two terms in the first line arise, respectively, from the field-induced shift of the minimum and from the interference phases of the harmonics. For a single harmonic they cancel identically. The local-derivative form also applies when the Fourier moments do not converge, as occurs for $S_4$ and $S_6$ in the zero-temperature diffusive example below. Thus, to leading order in the weak-texture limit, a finite cubic Josephson nonlinearity requires both an asymmetric spatial phase texture, $\kappa_3\neq0$, and more than one local Josephson harmonic.

\begin{figure}[t]
\includegraphics[width=\columnwidth]{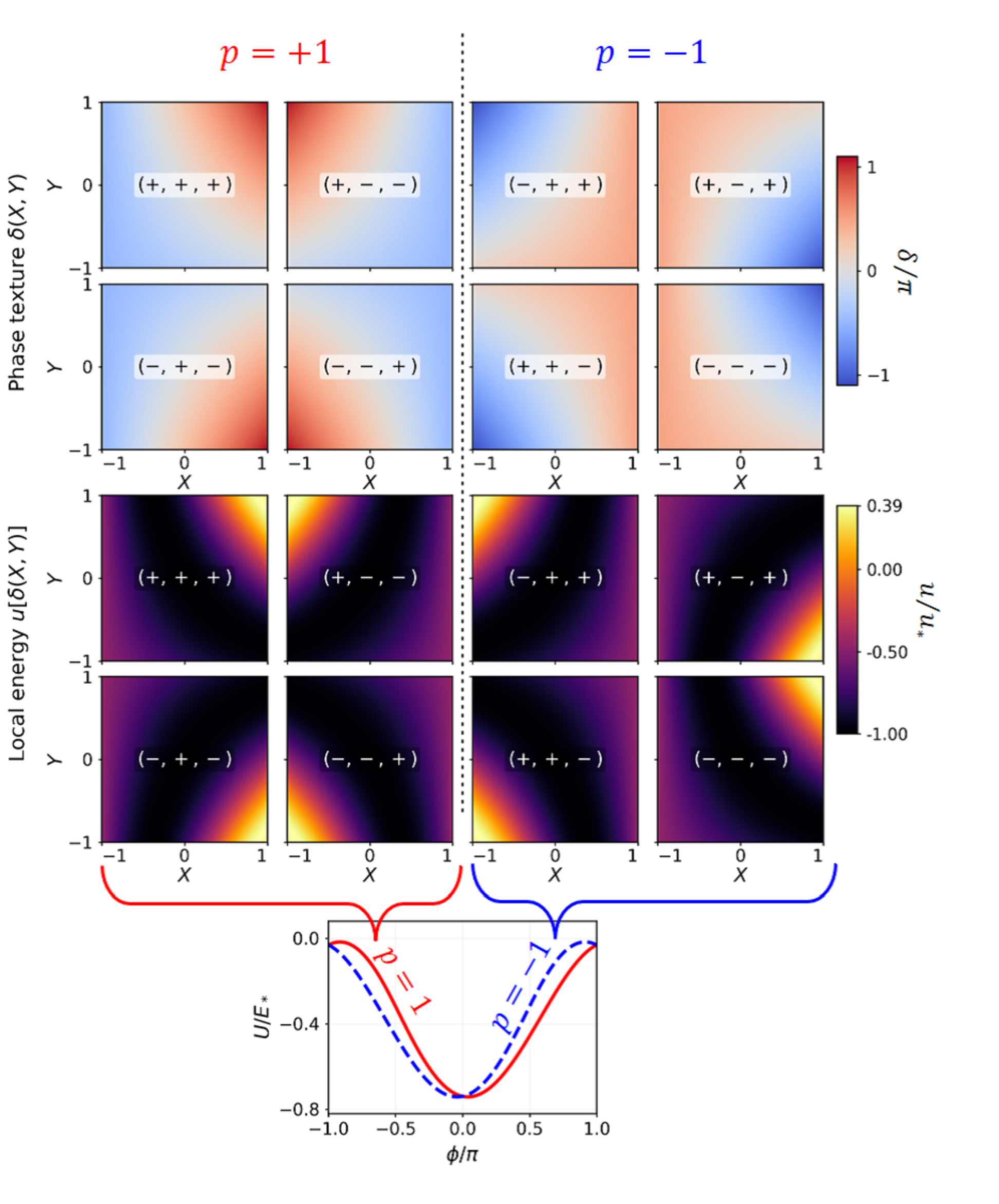}
\caption{
Magnetic-field-octant symmetry of the cross Josephson junction.
The upper and middle panels show the spatial phase texture
$\delta(X,Y)$ and local Josephson energy $u[\delta(X,Y)]$
at $\phi=0$, respectively, for the eight sign combinations
of $(\varphi_x,\varphi_y,\varphi_z)$.
The field magnitudes are fixed at
$(|\varphi_x|,|\varphi_y|,|\varphi_z|)/\pi=(0.30,0.50,0.30)$.
The left and right groups correspond to $p=+1$ and $p=-1$,
respectively, where $p=s_xs_ys_z$.
For a zero-temperature short diffusive SNS junction with transparent interfaces,
we use the zero-field current--phase relation
$I_0(\theta)=(\pi\Delta/eR_N)\cos(\theta/2)
\operatorname{artanh}[\sin(\theta/2)]$
~\cite{KulikOmelyanchuk1975,KulikOmelyanchuk1978,Golubov2004}, where $\Delta$ is the superconducting gap and $R_N$ the normal-state resistance of the whole junction.
The energy scales are $E_*=\Phi_0 I_*/(2\pi)$ and
$u_*=E_*/A_J$, where $I_*=\pi\Delta/(eR_N)$ and
$A_J=w_1w_2$, with the energy origin chosen such that
$u(0)=-u_*$.
The lower panel shows the spatially averaged potential
$U(\phi)$ for $p=+1$ (red solid line) and $p=-1$
(blue dashed line).
Reversing an odd number of field components maps
$U(\phi)$ onto $U(-\phi)$, whereas reversing an even
number leaves $U(\phi)$ unchanged.
Consequently, expansion coefficients about the corresponding minima
reverse sign at odd orders and remain unchanged at even orders.
}
\label{fig1}
\end{figure}

\textit{Example: Cross Josephson junction.---}
We now specialize to the cross junction, where the magnetic field produces a concrete two-dimensional phase texture and an exact field-reversal symmetry.
For $B_z$, the thin-strip fields $B_{c1}$ and $B_s$ provide reference scales for vortex-free operation~\cite{Kubo2026Bc1,Kubo2026Bs}.
We derive the spatial phase texture $\delta(x,y)$ entering the above equations. The upper strip extends along $y$ with width $w_1$ along $x$, and the lower strip extends along $x$ with width $w_2$ along $y$. 
Define the dimensionless coordinates $X=2x/w_1$ and $Y=2y/w_2$. The gauge-invariant Josephson phase difference is $\theta(x,y)=\chi_1(x,y,z_1)-\chi_2(x,y,z_2)-(2\pi/\Phi_0)\int_{z_2}^{z_1}A_z\,dz$, where $\chi_{1,2}$ are the condensate phases in the upper and lower electrodes. For a uniform field $\bm B=(B_x,B_y,B_z)$, we choose $\bm A=(-B_zy,0,B_xy-B_yx)$.
The London relation $\bm j_{1,2}=(\mu_0\lambda_{1,2}^2)^{-1}[(\Phi_0/2\pi)\nabla\chi_{1,2}-\bm A]$ determines the field-induced condensate phases. In the thin-electrode derivation we evaluate the phases at the electrode midplanes, approximate each strip as longitudinally infinite, and neglect end effects and Josephson-current redistribution. For the perpendicular-field response, longitudinal translation symmetry together with the insulating lateral boundary gives $j_{1x}=0$ in the upper strip, hence $\partial_x\chi_1=(2\pi/\Phi_0)A_x=-2\pi B_zy/\Phi_0$. Hence $\chi_1=\chi_{10}-2\pi B_zxy/\Phi_0$ in the absence of an imposed longitudinal current. Similarly, in the lower strip $j_{2y}=0$, and with $A_y=0$ we obtain $\chi_2=\chi_{20}$ in the absence of an imposed longitudinal current. The line integral of $A_z$ then gives the in-plane-field contribution proportional to the separation between the electrode midplanes.
For the in-plane-field contribution at finite electrode thickness, the geometric separation is replaced by an effective magnetic thickness determined by London screening; the perpendicular-field texture is retained in the weak-screening approximation. For a film of thickness $d_s$ centered at $z=0$, let $B_\parallel(z)$ denote the in-plane magnetic field and let $B_a=B_\parallel(\pm d_s/2)$ be its surface value.
The London equation then gives $B_\parallel(z)=B_a\cosh(z/\lambda)/\cosh[d_s/(2\lambda)]$. The field-weighted distance from the film center to the junction-facing surface is therefore $\int_0^{d_s/2}(B_\parallel/B_a)\,dz=\lambda\tanh[d_s/(2\lambda)]$. Adding the barrier and the contributions from the two electrodes gives $d_\parallel=d_j+\lambda_1\tanh[d_{s1}/(2\lambda_1)]+\lambda_2\tanh[d_{s2}/(2\lambda_2)]$.
Substituting these results into the gauge-invariant phase gives $\theta(x,y)=\phi+(2\pi/\Phi_0)[d_\parallel(B_yx-B_xy)-B_zxy]$, where $\phi=\chi_{10}-\chi_{20}$ is the spatially uniform circuit phase. In terms of $X$ and $Y$, this becomes $\theta(X,Y)=\phi+\delta(X,Y)$ with 
\begin{eqnarray}
\delta(X,Y)=\varphi_xY+\varphi_yX+\varphi_zXY, \label{delta_crossJJ}
\end{eqnarray}
where $\varphi_x=-\pi B_xd_\parallel w_2/\Phi_0$, $\varphi_y=\pi B_yd_\parallel w_1/\Phi_0$, and $\varphi_z=-\pi B_zw_1w_2/(2\Phi_0)$. The upper panels of Fig.~\ref{fig1} illustrate the resulting nonseparable phase texture generated by the $XY$ term.

Substituting the phase texture $\delta(X,Y)$ [Eq.~(\ref{delta_crossJJ})] into Eq.~(\ref{eq:Fm1}) gives the form factor for the cross JJ,
\begin{eqnarray}
F_m=\frac{1}{2}\int_{-1}^{1}dX\,
e^{im\varphi_yX}\,
\mathrm{sinc}[m(\varphi_x+\varphi_zX)].
\label{eq:Fm2}
\end{eqnarray}
All magnetic-field dependence is contained in the harmonic-dependent form factors $F_m$. As a simple check, for a single local harmonic one has $U=2A_J u_1|F_1|\cos(\phi+\alpha_1)$, and therefore $I=(2\pi/\Phi_0)dU/d\phi$ is sinusoidal with amplitude proportional to $|F_1|$. Hence the critical current is proportional to $|F_1|$.
If $B_z=0$ (and hence $\varphi_z=0$), $F_1$ factorizes as $F_1=\mathrm{sinc}(\varphi_x)\mathrm{sinc}(\varphi_y)$, recovering the familiar product of one-dimensional Fraunhofer factors. 
With multiple local harmonics, however, the phases of $F_m$ realize the mechanism discussed in the previous section, allowing the effective Josephson potential to become asymmetric even though the local energy remains reciprocal.

The general criterion derived above now becomes explicit for the cross junction. The harmonic-dependent phases $\alpha_m=\arg F_m$ are generated by the nonseparable phase texture, and a mismatch $\chi_m=\alpha_m-m\alpha_1$ that is nonzero modulo $\pi$ for an active harmonic signals the loss of a common inversion center when $u_1F_1\neq0$. In the weak-field limit, $|\varphi_x|$, $|\varphi_y|$, and $|\varphi_z|$ are small, and the phase texture $\delta(X,Y)$ is weak. For the cross junction, $\langle\delta\rangle=0$ and $\kappa_3=\langle\delta^3\rangle=(2/3)\varphi_x\varphi_y\varphi_z$. Using the weak-texture expansion derived above, $\alpha_m=-(m^3/9)\varphi_x\varphi_y\varphi_z+O(B^5)$ and hence $\chi_m=-[(m^3-m)/9]\varphi_x\varphi_y\varphi_z+O(B^5)$. Thus the leading mismatch for each fixed $m\geq2$ is nonzero when all three field components are present.
Using the general weak-texture result derived above, with $S_k=\sum_{m\geq1}u_m m^k$, we obtain
\begin{eqnarray}
c_3
&=&\frac{2A_J \varphi_x\varphi_y\varphi_z}{9}
\left(\frac{S_4^2}{S_2}-S_6\right) +O(B^5) \nonumber \\
&=&\frac{A_J\pi^3 d_\parallel^2 w_1^2 w_2^2}{9\Phi_0^3}
\left(\frac{S_4^2}{S_2}-S_6\right) B_xB_yB_z
+O(B^5).
\label{eq:c3general}
\end{eqnarray}
To this order, the cubic nonlinearity changes sign when any one of the three flux components is reversed and vanishes if any one of them is zero. For a single local harmonic, the term in parentheses vanishes identically, consistent with the general result that spatial interference alone cannot generate a cubic nonlinearity from a purely sinusoidal local junction. 
For the minimal case with only the first two harmonics,
\begin{eqnarray}
c_3
&=&-\frac{4A_J\pi^3d_\parallel^2w_1^2w_2^2}{\Phi_0^3}
\frac{u_1u_2}{u_1+4u_2}
B_xB_yB_z+O(B^5)
\nonumber\\
&\simeq&-\frac{4A_J\pi^3d_\parallel^2w_1^2w_2^2}{\Phi_0^3}
u_2 B_xB_yB_z , \label{eq:c3u2}
\end{eqnarray}
where the last form assumes $|u_2|\ll|u_1|$. Thus even a weak second harmonic produces a cubic response linear in both $u_2$ and $B_xB_yB_z$.

We examine how the form factor $F_m$ changes under magnetic-field reversal. Let $B_i\to s_iB_i$ ($i=x, y, z$) with $s_i=\pm1$ and $p=s_xs_ys_z$. Since $\varphi_i\propto B_i$, the coordinate changes $X\to ps_yX$ and $Y\to ps_xY$ in the spatial average give $F_m\to F_m$ for $p=+1$ and $F_m\to F_m^*$ for $p=-1$. Because all $u_m$ are real, $U(\phi)\to U(\phi)$ for $p=+1$ and $U(\phi)\to U(-\phi)$ for $p=-1$. Thus, reversing an even number of field components leaves the potential unchanged, whereas reversing an odd number reflects it about $\phi=0$. This result is exact within the stated phase-texture model. Consequently, about the corresponding minimum, even-order expansion coefficients are unchanged while odd-order ones reverse sign. 
In particular, $c_2$ is unchanged and $c_3$ changes sign, allowing the cubic interaction to be reversed without changing the linear resonance frequency.
For the zero-temperature short diffusive SNS example in Fig.~\ref{fig1}, evaluated from the full spatial average rather than the weak-field expansion, we obtain $c_2(B)/c_2(0)=0.6645$ and $c_3/c_2=-0.1577$ for the $(+,+,+)$ field octant. Reversing an odd number of field components preserves $c_2$ while changing $c_3/c_2$ to $+0.1577$.

\textit{Other examples.---}
The cross junction is not unique in realizing this mechanism. More generally, any extended reciprocal weak link with a spatial phase texture whose distribution lacks an inversion center can generate harmonic-dependent complex form factors. Possible realizations include non-centrosymmetric junction geometries, such as triangular or wedge-shaped weak links in a uniform field, and nominally symmetric junctions subjected to a spatially nonuniform magnetic field produced by local control lines or flux focusing. In all such cases, multiple local Josephson harmonics can convert the spatial phase asymmetry into an effective cubic interaction.


\textit{Quantum-circuit embedding and isospectral reversal.---}
We have so far focused on the cross junction as a concrete realization of the symmetry mechanism. We now return to the more general setting. Whenever an operation reflects the Josephson potential as $U(\phi)\to U(-\phi)$, we may ask how this transformation affects the corresponding quantum circuit. In particular, we ask whether the quantum energy levels remain unchanged when the odd nonlinearities change sign.

For a capacitively shunted junction, the Hamiltonian is~\cite{Koch2007}
\begin{eqnarray}
\hat H(n_g)=4E_C(\hat n-n_g)^2+U(\hat\phi),
\end{eqnarray}
where $E_C=e^2/(2C)$ is the charging energy, $C$ is the total capacitance across the junction, and $n_g=Q_g/(2e)$ is the dimensionless offset charge. We use periodic phase wavefunctions and conjugate variables with $\hat n=-i\partial_\phi$, formally $[\hat\phi,\hat n]=i$. Defining the Cooper-pair charge-translation operator $\hat T=e^{i\hat\phi}$ gives $\hat T\hat n\hat T^\dagger=\hat n-1$ and hence $\hat T\hat H(n_g)\hat T^\dagger=\hat H(n_g+1)$. We denote unitary equivalence by $\sim$, so that $\hat H(n_g)\sim\hat H(n_g+1)$. If $\hat H(n_g)|\psi_j\rangle=E_j|\psi_j\rangle$, then $\hat H(n_g+1)\hat T|\psi_j\rangle=E_j\hat T|\psi_j\rangle$, showing explicitly that unitarily equivalent Hamiltonians have identical eigenvalues. The spectrum is therefore periodic under $n_g\to n_g+1$.

For the reflection $U_s(\phi)=U(-\phi)$, realized by odd field reversal ($p=-1$) in the cross junction, define the unitary phase-parity operator by $(\hat\Pi\psi)(\phi)=\psi(-\phi)$. Since $\hat n=-i\partial_\phi$ in the phase representation, it follows that
$\hat\Pi\hat\phi\hat\Pi^\dagger=-\hat\phi$
and
$\hat\Pi\hat n\hat\Pi^\dagger=-\hat n$.
Therefore,
$\hat H_s(n_g)=\hat\Pi\hat H(-n_g)\hat\Pi^\dagger$.
Combining this relation with the unit-periodicity in $n_g$ gives
$\hat H_s(n_g)\sim\hat H(-n_g+N)$ for any $N\in\mathbb Z$,
where $\sim$ denotes unitary equivalence.
At $n_g=0$, $\hat H_s(0)\sim\hat H(0)$. At $n_g=1/2$ modulo one, $-n_g$ is equivalent to $n_g$ by the unit-periodicity. Thus, integer and half-integer offset charges admit a unitary implementation of the reflection.
More generally, let $(\mathcal K\psi)(\phi)=\psi^*(\phi)$ and define the antiunitary operator $\mathcal A=\hat\Pi\mathcal K$. It satisfies $\mathcal A\hat\phi\mathcal A^{-1}=-\hat\phi$ and $\mathcal A\hat n\mathcal A^{-1}=\hat n$, and preserves the periodic boundary conditions. Hence
\begin{eqnarray}
\hat H_s(n_g)=\mathcal A\hat H(n_g)\mathcal A^{-1}.
\label{eq:antiunitary}
\end{eqnarray}
The reflected potentials are therefore isospectral for arbitrary $n_g$ in this capacitively shunted model, while the odd-order potential coefficients about corresponding minima reverse sign.

The embedding extension requires a symmetry of the complete static Hamiltonian, including biases and boundary conditions. For example, $\mathcal A$ acting on all coordinates suffices for real quadratic capacitive energy and reflection-even inductive energy, provided boundary conditions are preserved. The mapped equilibria then have identical Hessians and linear spectra, whereas odd junction terms change sign. Reciprocity alone is insufficient.


\textit{Three-wave-mixing.---}
We now consider how the isospectral sign reversal manifests itself in three-wave mixing. We embed the junction in a linear electromagnetic environment described by $\hat H_{\rm em}$. The total Hamiltonian is
\begin{eqnarray}
\hat H_{\rm tot}
=
\hat H_{\rm em}
+4E_C(\hat n-n_g)^2
+U(\hat\phi).
\end{eqnarray}
Here $\hat H_{\rm em}$ contains the remaining quadratic circuit terms and their couplings to the junction coordinates, with the junction capacitance counted only once.
When the junction is embedded in this environment, the equilibrium phase across the junction can in general differ from the minimum $\phi_0$ of the isolated junction. We therefore denote the junction phase at the equilibrium of the complete circuit by $\bar{\phi}$ and define the fluctuation $\hat q:=\hat\phi-\bar{\phi}$.
Expanding the junction potential about the equilibrium
of the complete circuit gives
$U(\bar{\phi}+\hat q)
=U(\bar{\phi})+U'(\bar{\phi})\hat q
+\bar c_2\hat q^2/2
+\sum_{m\ge3}\bar c_m\hat q^m/m!$,
where $\bar c_m:=U^{(m)}(\bar{\phi})$.
Although $U'(\bar{\phi})$ need not vanish, its linear
contribution cancels that from the electromagnetic
environment because the complete circuit is at equilibrium.
Apart from the total equilibrium energy, the total Hamiltonian can therefore be written as
\begin{eqnarray}
&&\hat H_{\rm tot}
=\hat H_{\rm lin}+\sum_{m\ge3}\hat V_m,\\
&&\hat H_{\rm lin}
=\hat H_{\rm em}^{(2)}
+4E_C(\hat n-n_g)^2
+\frac{\bar c_2}{2}\hat q^2,\\
&&\hat V_m
=\bar c_m\frac{\hat q^m}{m!},
\end{eqnarray}
where $\hat H_{\rm em}^{(2)}$ denotes the quadratic expansion of the linear electromagnetic environment about the same equilibrium.
Within the small-fluctuation approximation about a stable equilibrium, diagonalizing the quadratic circuit gives independent harmonic normal modes~\cite{Nigg2012}. If only one normal mode of the complete linearized circuit is retained, the linear Hamiltonian can be written as
$\hat H_{\rm lin}=\hbar\omega(\hat a^\dagger\hat a+1/2)$.
The corresponding contribution to the junction phase fluctuation is
$\hat q=q_{\rm zpf}(\hat a+\hat a^\dagger)$,
where $q_{\rm zpf}$ is the zero-point phase amplitude across the junction.
For a multimode environment, these expressions generalize to
$\hat H_{\rm lin}=\sum_j\hbar\omega_j(\hat a_j^\dagger\hat a_j+1/2)$
and
$\hat q=\sum_j q_{{\rm zpf},j}(\hat a_j+\hat a_j^\dagger)$.

To illustrate the cubic interaction, consider three linear resonator modes, labeled $1$, $2$, and $3$, with mode $3$ serving as the pump and modes $1$ and $2$ as the signal and idler. After diagonalizing the complete quadratic Hamiltonian, these modes have frequencies $\omega_1$, $\omega_2$, and $\omega_3$. 
Substituting the normal-mode expansion, $\hat q=\sum_{\mu=1}^{3}q_\mu^{\rm zpf}(\hat a_\mu+\hat a_\mu^\dagger)$, into the cubic part $\hat{V}_3$, the terms containing one operator from each of the three modes appear with a factor $3!=6$, giving
$\bar c_3 q_1^{\rm zpf}q_2^{\rm zpf}q_3^{\rm zpf}
(\hat a_1+\hat a_1^\dagger)
(\hat a_2+\hat a_2^\dagger)
(\hat a_3+\hat a_3^\dagger)$.
In the interaction picture, the term
$\hat a_1^\dagger\hat a_2^\dagger\hat a_3$
acquires the phase factor
$e^{i(\omega_1+\omega_2-\omega_3)t}$,
and therefore becomes resonant when
$\omega_3\simeq\omega_1+\omega_2$.
Under this condition, the slowly rotating terms are
$\hat a_1^\dagger\hat a_2^\dagger\hat a_3$
and
$\hat a_1\hat a_2\hat a_3^\dagger$.
Keeping only these terms gives the resonant three-wave-mixing contribution to the nonlinear Hamiltonian,
\begin{eqnarray}
\hat V_3 \ni \hat H_{3\mathrm{wm}}
=\hbar g 
\left(
\hat a_1^\dagger\hat a_2^\dagger\hat a_3+
\hat a_1\hat a_2\hat a_3^\dagger
\right),
\label{eq:3wm}
\end{eqnarray}
with
$g:=(\bar c_3/\hbar)q_1^{\rm zpf}q_2^{\rm zpf}q_3^{\rm zpf}$.
For a reflection-related embedding with fixed real mode conventions, the same $q_\mu^{\rm zpf}$ can be used in the two configurations. The sign operation then sends $\bar c_3\to-\bar c_3$ and $g\to-g$ without changing the linear normal-mode frequencies.

To estimate the size of this interaction, consider the zero-temperature short diffusive SNS operating point in Fig.~\ref{fig1}, for which the spatially averaged potential gives $c_2(B)/c_2(0)=0.6645$ and $c_3/c_2=-0.1577$. Taking a single-mode reference circuit at $n_g=0$ with $q_{\rm zpf}=(2E_C/c_2)^{1/4}=0.10$ fixes $E_C/E_*=1.6612\times10^{-5}$. Charge-basis diagonalization of the full periodic potential gives $E_*/h\simeq904.6$ GHz when the first transition is set to $6$ GHz. With this scale, we obtain the cubic coupling $g_3=c_3q_{\rm zpf}^3/(6\hbar)$, giving $g_3/2\pi\simeq-7.90$ MHz. Keeping this junction energy scale and taking three modes with $q_1^{\rm zpf}=q_2^{\rm zpf}=q_3^{\rm zpf}=0.03$, assuming a negligible equilibrium shift upon embedding, Eq.~(\ref{eq:3wm}) gives $g/2\pi\simeq-1.28$ MHz. These values illustrate Hamiltonian coupling scales, not loss or coherence predictions; a complete device requires a specific multimode electromagnetic design.


\textit{Phase-sensitive test of local Josephson harmonics.---}
The cubic response can also distinguish genuine local
Josephson harmonics from nonlinearities that might otherwise
be attributed to the surrounding linear circuit.
For a locally sinusoidal junction with a common linear series inductance, the energy is $V(\phi,\psi)=E_L(\phi-\psi)^2/2-E_1\cos(\psi+\alpha_1)$, where $E_L,E_1>0$ and $\psi$ is the junction phase. Shifting both phase coordinates by $\alpha_1$ makes $V$ invariant under their simultaneous reflection. Eliminating the internal coordinate therefore leaves an effective potential even about the unfrustrated equilibrium. More general embeddings require this same common reflection symmetry and a reflection-invariant equilibrium; reciprocity alone is insufficient. In particular, an inductive shunt fixing a different phase origin is not covered by this null test.
In contrast, a genuine local second harmonic $u_2$
produces the finite response in Eq.~(\ref{eq:c3u2}).

Because $\bar c_3$ is not measured directly, we convert the symmetry test into a phase-sensitive microwave signal. We consider the degenerate counterpart of the three-mode process discussed above, in which the two low-frequency quanta occupy the same normal mode $1$, with $\omega_3\simeq2\omega_1$, rather than two distinct modes of equal frequency. The relevant phase fluctuation is then
$\hat q=q_1^{\rm zpf}(\hat a_1+\hat a_1^\dagger)+q_3^{\rm zpf}(\hat a_3+\hat a_3^\dagger)$.
Expanding $\hat V_3=(\bar c_3/3!)\hat q^3$, the term containing two mode-$1$ operators and one mode-$3$ operator carries the combinatorial factor $3$, yielding the factor $1/2$ below. The resonant part of the cubic interaction is therefore
\begin{eqnarray}
\hat V_3 \ni \hat H_{113}
=
\hbar g_{113}
\left(
\hat a_1^2\hat a_3^\dagger+
\hat a_1^{\dagger 2}\hat a_3
\right),
\label{eq:113}
\end{eqnarray}
with
$g_{113}:=\bar c_3(q_1^{\rm zpf})^2q_3^{\rm zpf}/(2\hbar)$.
In the weak-conversion, undepleted-drive limit, driving mode $1$ with rotating-frame amplitude $\alpha=\langle\hat a_1\rangle$ generates a steady amplitude $\beta_3=-ig_{113}\alpha^2/(\Gamma_3/2+i\Delta_3)$ in mode $3$, where $\Gamma_3$ is its energy decay rate and $\Delta_3=\omega_3-2\omega_d$ for drive frequency $\omega_d$; we use the standard linear input--output convention~\cite{Blais2021}. With the drive and detection phase references, mode participations, and linear input--output response fixed, a sign-reversal operation (e.g., odd field reversal in the cross junction) sends $g_{113}\to-g_{113}$ and hence $\beta_3\to-\beta_3$. Intensity alone does not detect this sign. Subtracting the complex signals in the two symmetry-related configurations isolates the sign-changing contribution while canceling unchanged backgrounds; its attribution to local harmonics requires the common-reflection condition stated above.

The sign relation is not tied to the cross geometry: it applies to any implementation with the full-Hamiltonian reflection and measurement conditions specified above. In the imposed-field cross-junction model, odd field reversal supplies the required junction-potential reflection.

The same symmetry criterion also connects directly to the Josephson diode effect, as studied in higher-harmonic interferometers~\cite{FominovMikhailov2022,SeoaneSouto2022}. Define $I_c^+=\max_\phi I(\phi)$, $I_c^-=\min_\phi I(\phi)<0$, and the diode efficiency $\eta=(I_c^+-|I_c^-|)/(I_c^++|I_c^-|)$. An inversion-symmetric potential gives $\eta=0$. For the cross junction in the weak-field limit, spatial averaging of an arbitrary zero-field current--phase relation $I_0(\phi)$ gives
\begin{eqnarray}
\eta
=
\frac{\pi^3 d_\parallel^2 w_1^2 w_2^2}{18\Phi_0^3}
\frac{I_0'''(\phi_*)}{I_0(\phi_*)}
B_xB_yB_z
+O(B^5),
\label{eq:diode_general}
\end{eqnarray}
where $\phi_*$ is assumed to be the unique, nondegenerate positive critical-current phase, $I_0'(\phi_*)=0$ and $I_0''(\phi_*)<0$, with $I_0$ smooth near $\pm\phi_*$. Here $O(B^5)$ denotes total degree five and higher in the field components. Thus all three field components are required at leading order, and reversing any one of them reverses the diode polarity. For the minimal two-harmonic case, $I_0=I_1\sin\phi+I_2\sin2\phi$ with $I_1>0$ and $|I_2|\ll I_1$, this reduces to $\eta\simeq [2\pi^3d_\parallel^2w_1^2w_2^2/(3\Phi_0^3)](u_2/u_1)B_xB_yB_z$. The diode response is a complementary dc signature, not a direct measure of $c_3$: for example, a current--phase relation containing only odd harmonics satisfies $I(\phi+\pi)=-I(\phi)$ and has $\eta=0$ even when $c_3\neq0$.

\textit{Conclusion.---}
Spatial interference can change the inversion symmetry of an effective Josephson potential even when the local energy remains inversion symmetric. Different local harmonics acquire incompatible interference phases, allowing odd nonlinearities that cannot be removed by a common phase shift. A homogeneous cross junction realizes this mechanism, yielding $c_3\propto B_xB_yB_z$ at weak field and an exact potential reflection under odd field reversal. This operation reverses odd nonlinearities while preserving the complete spectrum of a capacitively shunted junction at arbitrary offset charge. The same mechanism can produce a Josephson diode response and, under suitable circuit symmetry conditions, provide a phase-sensitive test of genuine local harmonics. Thus spatial interference offers a means to engineer not only the strength but also the symmetry of nonlinear Josephson interactions.



\end{document}